\documentclass[pra,aps,twocolumn,superscriptaddress,showpacs,amsmath,amssymb,floatfix]{revtex4}

\usepackage{graphicx}
 
\newcommand{\be}{\begin{equation}}
\newcommand{\ee}{\end{equation}}

\newcommand{\kk}{{\mathbf k}}
\newcommand{\eq}[1]{(\ref{#1})}
\newcommand{\mume}{\mu \rm m}
\newcommand{\im}{\textrm{Im}}
\newcommand{\re}{\textrm{Re}}
\newcommand{\Psihd}{{\hat \Psi}^\dagger}
\newcommand{\Psih}{{\hat \Psi}}
\newcommand{\Ehd}{{\hat E}^\dagger}
\newcommand{\Eh}{{\hat E}}

\begin{document}
\title{The Goldstone mode of planar optical parametric oscillators}
\author{Michiel Wouters}
\affiliation{BEC-CNR-INFM and Dipartimento di Fisica, Universit\`a di Trento, 
I-38050 Povo, Italy}
\affiliation{TFVS, Universiteit Antwerpen, Universiteitsplein 1,
2610 Antwerpen, Belgium}
\author{Iacopo Carusotto}
\affiliation{BEC-CNR-INFM and Dipartimento di Fisica, Universit\`a di Trento, 
I-38050 Povo, Italy}
\begin{abstract}
We propose an experimental setup to probe the low-lying excitation modes of
a parametrically oscillating planar cavity, in particular the soft
Goldstone mode which appears as a consequence of the spontaneously
broken $U(1)$ symmetry of signal/idler phase rotations.
Differently from the case of thermodynamical equilibrium, the
Goldstone mode of such a driven-dissipative system is an overdamped
mode, whose linewidth goes to zero in the long-wavelength limit.
When the phase of the signal/idler emission is pinned 
by an additional laser beam in the vicinity of the signal
emission, the $U(1)$ symmetry is explicitely broken and a gap opens in
the Goldstone mode dispersion. This results in a dramatic broadening
of the response to the probe.
Quantitative predictions are given for the case of semiconductor
planar cavities in the strong exciton-photon coupling regime. 
\end{abstract}
\maketitle

\section{Introduction}

A central concept of the modern theory of phase transitions
in systems at thermodynamical equilibrium is the so-called Goldstone
mode, which appears as a consequence of the spontaneous breaking of a
continuous symmetry~\cite{goldstone-stat-mech}.
Unless the system has long-range interactions \cite{anderson}, this mode has the general
property of being a soft mode, whose frequency dispersion $\omega_G(k)$
tends to zero in the long-wavelength limit $k\rightarrow 0$.
The
Goldstone mode corresponds in fact to a spatially slow twist of the
order parameter, whose energy cost tends to zero in the
long-wavelength limit.
Among the most celebrated examples of Goldstone modes in
condensed matter physics we can mention the zero-sound mode of
superfluid Helium 4 or dilute Bose-Einstein
condensates and the magnon excitations in ferromagnets.
Zero-sound is related to the spontaneous breaking of the $U(1)$ gauge
symmetry of the quantum Bose field below
the Bose-Einstein condensation
temperature~\cite{goldstone-stat-mech,forster,pines-nozieres}, while the magnon branch is
related to the spontaneous breaking of the rotational symmetry of the
magnetic moment orientation below the Curie temperature~\cite{magnon}.

The concept of a Goldstone mode plays an important role also in
the physics of nonlinear dynamical systems far from
thermodynamical equilibrium~\cite{cross},
whose stationary state is not determined by 
a thermodynamical equilibrium condition, but is rather the
result of a dynamical equilibrium between an external driving force
and the dissipation.
In many examples of such driven-dissipative systems a 
homogeneous state develops a non-trivial spatiotemporal pattern 
when the driving force exceeds a critical value.
The most famous example of this pattern formation behaviour is perhaps
the regular periodic  
arrangement of B\'enard cells that appears in heat convection through a viscous
fluid when a sufficiently large temperature gradient (driving force)
is applied in the vertical direction so to exceed the braking effect
of viscosity (dissipation)~\cite{benard}.
The continuous translational symmetry of the
initially homogeneous system is reduced to the
discrete symmetry of the periodic pattern of B\'enard cells.
As a consequence of the spontaneously broken symmetry, a neutral mode
of vanishing frequency and damping rate appears in the linear
stability analysis around the stationary state of the system, a mode
which corresponds to the rigid translation of the whole roll pattern
in space. This neutral mode is the non-equilibrium counterpart of the
Goldstone mode of equilibrium statistical mechanics.

Another well celebrated example of pattern formation
takes place in optical parametric oscillation (OPO) in planar
cavities~\cite{OPO_exp}. 
Differently from the more usual case of spherical mirror cavities with
a discrete and well-spaced set of modes, 
planar cavities dispose of a continuum of modes in which the
parametric emission can take place, so that the light field has
a rich spatial dynamics~\cite{planar_OPO_2}.
In the OPO state, a periodic spatial pattern is indeed created  in the
cavity plane by interference of the pump, signal and idler fields.
In addition to their interest from the point of view of fundamental
physics, technological applications of planar OPOs have also been
actively investigated in view of using them as flexible light sources 
in new frequency ranges, or even as sources of entangled photons for
quantum  cryptography~\cite{quantum_crypt}.

In recent years, a great deal of theoretical and experimental
  activity has concerned planar
semiconductor microcavities with quantum well excitonic transitions
strongly coupled to the cavity mode~\cite{review1,review2}. 
At linear regime, the elementary excitations of these systems are
cavity-polaritons, i.e. a superposition of a cavity photon and a
  quantum well exciton. 
Cavity-polaritons  have the interesting property of combining the
extremely strong Kerr nonlinearity due to the excitonic component to
  a peculiar dispersion relation which allows for easy and robust
  phase-matching of the optical parametric process.
In this way, large values of gain have been observed in parametric
amplification~\cite{OPA_mcav}, as well as parametric oscillation at
low pump intensity values~\cite{stevenson,houdre}.  
Extensive studies of the parametric oscillation process in
semiconductor planar microcavities have verified the
coherent nature of the parametric emission above threshold by
observing a frequency narrowing of the emission~\cite{stevenson}, the
absence of dispersion~\cite{houdre}, as well as the long-range spatial
coherence~\cite{baas-coh}.

As it happened for atomic Bose-Einstein condensates~\cite{book}, 
a good deal of information on the state of the polariton system can be
obtained by looking at the elementary excitations of the system
around its dynamical equilibrium state.
The elementary excitations around the pump-only state below the
threshold for parametric oscillation have been theoretically
investigated in~\cite{superfl}, and predictions have been put forward
for polariton superfluidity effects.
In the present paper, we shall investigate the elementary excitation
spectrum around the parametric oscillation steady state above
threshold.
Since a continuous $U(1)$ symmetry related to the signal/idler phases is
spontaneously broken, a Goldstone mode has to appear in the
excitation  spectrum of the planar system, whose frequency and damping
rates go to zero in the $k\rightarrow 0$ long-wavelength limit.
Although its existence has been recently mentioned~\cite{zambrini} and its
role in the destruction of long-range coherence in 1D systems
discussed~\cite{zambrini,coherence}, no experiments have
been so far performed nor proposed which are able to directly observe
the Goldstone mode of optical parametric oscillators.
Pioneering work on the elementary
excitation spectrum around a parametrically oscillating state was
reported in Ref.\cite{ciuti-offbranch} both in the presence of a laser
beam driving the signal mode, and in a pump-only configuration. 
The emission was experimentally observed and its spectrum compared
to a calculation of the elementary excitation dispersion.
However, no specific attention was paid in the experiment to the
region where the Goldstone mode is expected to appear, nor any mention
made to it in the theoretical analysis. The same with other recent theoretical
papers that address the excitation spectrum around a parametrically
oscillating state of a semiconductor microcavity~\cite{savona,whittaker2}.

The main point of the present paper is to give a complete account
  of the physics of the Goldstone mode of a planar optical parametric   
oscillator above threshold, and to propose a way of
  probing it with an additional laser beam.
In Sec.\ref{sec:model} we present the physical system and the
theoretical framework used for its description. 
In Sec.\ref{sec:OPO} we study the stationary state of parametric
oscillation above threshold and we introduce the concept of
spontaneous breaking of the $U(1)$ signal/idler phase symmetry. 
The spectrum of the elementary excitations around the stationary
state is calculated and discussed in the following Sec.\ref{sec:lin},
and then used to evaluate the linear response of the system to a weak
probe beam at angles close to the signal emission: the Goldstone mode
is shown to give a strong and narrow peak at low frequencies,
whose linewidth tends to zero as the direction of the probe beam is 
brought closer to the signal emission one.
The consequences of the presence of an additional signal laser
field at exactly the same wavevector as the signal emission are
investigated in Sec. \ref{sec:destroy}.
From the full equations of motion, one finds that the phase of
  the signal emission results in this case pinned to the one of the
  signal beam so that the $U(1)$ symmetry is explicitely broken.
No Goldstone mode is therefore present any longer and a gap
opens up in the dispersion of the elementary excitations.
As a consequence, a dramatic broadening of the peak corresponding to
the Goldstone mode is observed in the response spectrum to the probe.
This phenomenology is the non-equilibrium analog to what happens in a
ferromagnet when an external magnetic field is applied to the system
to break the rotational symmetry. 
In this case, the orientation of the magnetization is fixed by the
applied field and a gap opens up in the magnon
dispersion~\cite{magnon}.

The two last section are devoted to the analysis of issues of
experimental relevance.
The consequences of a possible frequency mismatch of the applied
signal beam from the natural frequency of the signal emission are
investigated in Sec. \ref{sec:mismatch}, while the effect of the
spatial inhomogeneities 
due to the finite pump spot are addressed in Sec.\ref{sec:finite}.
Both these issues are shown not to affect the properties of the
Goldstone mode discussed in the previous sections.
Conclusions are drawn in Sec.\ref{sec:Conclu}.
Three appendices are devoted to a brief summary of the analytical
  equations definining the stationary state of the homogeneous system,
to the analytical properties of the Goldstone mode frequency around
$k=0$, and to the comparison  with the case of OPOs based on a
  $\chi^{(2)}$ second-order 
  nonlinearity.

\section{Physical system and theoretical model \label{sec:model}}

\begin{figure}
\begin{center}
\includegraphics[width=\columnwidth,angle=0,clip]{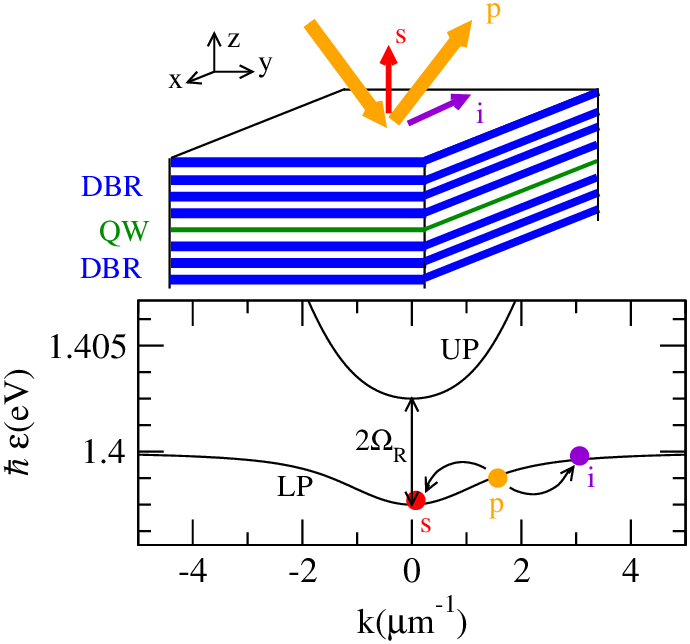}
\end{center}
\caption{\label{fig:sketch}
Upper panel: sketch of the microcavity system and the parametric
process under consideration.
Lower panel: Lower polariton (LP) and upper polariton (UP) 
dispersion. The cavity photon dispersion is 
$\omega_{C}(k)=\omega_{C}^0\,\sqrt{1+{k^2}/{k_z^2}}$
  with
$\hbar\omega^0_C=1.4$~eV and $k_z=20\,\mu\rm m^{-1}$.
The exciton dispersion is flat and resonant with the $k=0$ cavity mode
$\omega_X=\omega_C^0$. The exciton-photon Rabi coupling is
$\hbar\Omega_R=2.5$~meV. Dots indicate the signal, pump and
  idler modes.
}  
\end{figure}

The physical system we are considering consists of a planar
semiconductor microcavity containing a quantum well with an excitonic
transition strongly coupled to the cavity mode.
A sketch is shown in Fig.\ref{fig:sketch}.
The elementary excitations of this system are
cavity-polaritons, i.e. coherent superpositions of
cavity photons and excitons, which satisfy the Bose statistics. 
The photonic component is essential to couple to the external incident
radiation, while the excitonic component provides the exciton-exciton
collisional interactions which are responsible for the parametric process.

Given the translational invariance of the system along the cavity
plane, the in-plane wave vector $\kk$ is a good quantum number and the
polaritonic dispersion can be studied as a function of $\kk$.
As shown in Fig.\ref{fig:sketch}b, two polaritonic branches exist in
the polariton spectrum, split by twice the Rabi frequency $\Omega_R$
of the exciton-photon coupling~\cite{review1,review2}.
A simple, but so far accurate model for interactions is based on a
repulsive exciton-exciton two-body contact interaction of strength
$g$. 

In the following we will focus our attention on the lower polariton
branch only, which is more protected from loss and decoherence
processes. Its dispersion will be denoted as $\epsilon(k)$.
In order to justify the neglection of the upper polariton, one has to
check that the Rabi splitting $\Omega_R$ is much larger than 
both the detuning of the incident laser (of frequency
$\omega_p$ and wave vector $k_p$), and the nonlinear shift
$\epsilon_{mf}\lesssim g\,n_{LP}$ of the polariton mode ($n_{LP}$ is
here the polariton density).
These condition are actually well satisfied in current
experiments.

At the mean-field level, the dynamics of the polaritonic field can be
described by a polaritonic Gross-Pitaevskii
equation~\cite{ciuti_review,superfl}, whose 
$k$-space form reads:  
\begin{multline}
i \frac{d}{dt}\psi_{LP}(k) =\left[\epsilon(k) 
-i\frac{\gamma}{2}\right] \psi_{LP}(k) 
+F_{p}(k)\,\, e^{-i\omega_{p}t}  \\
+  \sum_{q_1,q_2} g_{k,q_1,q_2}\, \psi^*_{LP}(q_1+q_2-k)
\, \psi_{LP}(q_1)\, \psi_{LP}(q_2).
\label{eq_mot}
\end{multline}
$F_p(k)$ is here the amplitude of the incident pump laser field,
which is assumed to be continuous wave and monochromatic at $\omega_p$. 
Unless otherwise specified (as e.g. in Sec.\ref{sec:finite}), a
plane-wave at $k_p$ is taken for its spatial profile $F_p(k)=F_p
\delta_{k,k_p}$.
The damping rate $\gamma$ is the sum of the contribution of radiative
and non-radiative losses. It has been taken for simplicity as
momentum-independent. 
The momentum-dependent nonlinear interaction strength for polaritons
$g_{k,q_1,q_2}$ is defined in terms of the Hopfield coefficient $U_X$
quantifying the excitonic content of the
polariton~\cite{ciuti_review} as:
\begin{equation}
g_{k,q_1,q_2}=g\, U^*_X(k)\, U^*_X(q_1+q_2-k)\, U_X(q_1)\, U_X(q_2),
\end{equation}
Note that the same equation \eq{eq_mot} can be used to describe the
photon dynamics in different systems, e.g. metallic mirror planar
cavities containing a Kerr nonlinear medium~\cite{butcher}.
In this case, no excitonic field is present and the polaritonic field
$\Psih_{LP}$ reduces to the bare e.m. field $\Eh$.
The nonlinear coupling constant $g$ is then provided by the
Kerr optical nonlinearity of the cavity material (of dielectric
constant $\epsilon_{\rm lin}$ and thickness $d$), and is proportional
to its third-order nonlinear polarizability $\chi^{(3)}$: 
\begin{equation}
\hbar g=C \chi^{(3)}(\hbar\omega_0)^2/(d \epsilon_{\rm lin}^2),
\end{equation}
where $C$ is a numerical
factor of order one that takes into account the boundary conditions at
the cavity mirrors. 

Quantitatively, one however should keep in mind that the nonlinear coupling
constant $g$ for semiconductor microcavities in the strong
exciton-photon coupling regime $\hbar g\approx
5\times10^{-6}\,\textrm{eV}\, \mu \textrm{m}^2 $  
is orders of magnitude larger than the one that is obtained in
standard transparent media for nonlinear optics. 
Organic materials specifically suited for nonlinear optical
  applications have in fact $\chi^{(3)}$ up to about $10^{-9}\,
  \textrm{erg}^{-1}\,\textrm{cm}^{3}$~\cite{cont-nonlin-opt}, which
  yields values for the 
nonlinear coupling constant of the order of $\hbar g=5 \times 10^{-9}\,
\textrm{eV}\, \mu \textrm{m}^2$. Conventional inorganic materials have even weaker
susceptibilities.

\section{Parametric oscillation and spontaneous symmetry breaking}
\label{sec:OPO}
Among the many processes described by the motion equation \eq{eq_mot}
in the different regimes, the focus of the present paper 
will be concentrated on parametric process in which two polaritons in
the pump mode (wave vector $k_p$ and frequency 
$\omega_p$) collide and are converted into a pair of signal and
idler polaritons, of wave vectors respectively $k_s$ and
$k_i=2k_p-k_s$ and frequencies $\omega_s$ and
$\omega_i=2\omega_p-\omega_s$.
As shown in Fig.\ref{fig:sketch}, the peculiar polaritonic dispersion
allows for the parametric process 
to occur in a triply-resonant way, i.e. with all the $\omega_{p,s,i}$
close to resonance with the free polariton energy
$\epsilon(k_{p,s,i})$. 
Such a process is described by a term in the nonlinear interaction
Hamiltonian of the form:
\begin{multline}
H_{\rm int}=g_{k_p,k_s,k_i}\,\Big[\Psihd_{LP}(k_p)\,\Psihd_{LP}(k_p)\,
\Psih_{LP}(k_i)\,\Psih_{LP}(k_s)+\\ +{\textrm h.c.}\Big],
\label{par_ham}
\end{multline}
where $\Psih_{LP}$ is the polariton quantum field operator.

It is important to remind that OPO operation can also make use of a
different nonlinear process, in which a single pump photon is
split into a pair of signal/idler photons. The nonlinear
susceptibility involved in this parametric downconversion process
is the second-order one $\chi^{(2)}$.
Before the advent of semiconductor microcavities in the strong coupling
regime, most of the existing literature on OPOs was indeed based on
such parametric down conversion processes~\cite{planar_OPO_2}. 
Although our attention will be in the following concentrated on the
specific case of semiconductor microcavities, the concept of the
Goldstone mode extends without almost any change to the other case. A
brief discussion of these issues is given in Appendix~\ref{app:chi2}.

\begin{figure}[tb]
\begin{center}
\includegraphics[width=\columnwidth,angle=0,clip]{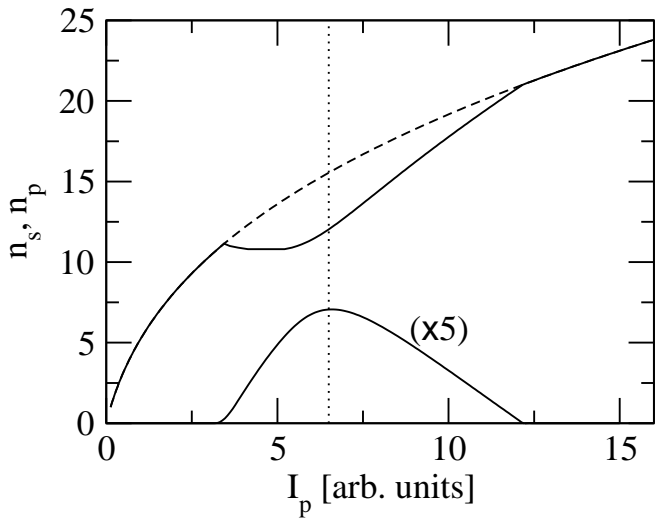}
\caption{
Pump and signal intensity as a function of the pump power for
$k_s=0.075 \mu \rm m^{-1}$. 
Pump parameters $k_p=1.57 \mu \rm m^{-1}$ and $\omega_p=1.39875$.
Cavity and exciton parameters correspond to the LP dispersion
  shown in figure 
\ref{fig:sketch}b. 
}
\label{fig:sp}
\end{center}
\end{figure} 

In order to study the parametric process the following ansatz can
  be used for $\psi_{LP}(x,t)$:
\begin{multline}
\psi_{LP}(x,t) =P\, e^{i(k_p x-\omega_{p}t)}+
S\, e^{i(k_s x-\omega_{s}t)}+\\
+I\, e^{i(k_i x- \omega_i t)},
\label{psi0}
\end{multline}%
where $k_s$ is considered here as a given parameter. 
A detailed analysis of its behaviour as a function of the geometry
and system parameters is postponed to a forthcoming
publication~\cite{pattern}.

This ansatz allows for finite amplitudes $P,S,I$ in respectively the
pump, signal and idler modes, while all other modes are assumed to remain empty.
The possible occupation of other modes via multiple scattering
  processes, as observed in~\cite{ciuti-offbranch,tarta} is therefore
  not taken into account by our model.
This approximation is well justified by the fact that the (small)
polariton population in these modes does not affect the physics
under examination but can only give quantitatively small corrections. 
The values of $P,S,I$ in the stationary state, as well as the
parametric oscillation frequency $\omega_s$ are obtained by inserting
this ansatz in the equation of motion \eq{eq_mot}.
Details of the equations are given in Appendix~\ref{app:calculation}.

The pump and signal intensities in the stationary state are plotted in
Fig.\ref{fig:sp} 
as a function of the pump intensity $I_p=|F_p|^2$ for a pump frequency
$\omega_p<\epsilon(k_p)$, in which case the pump-only dynamics
corresponds to optical  limiting~\cite{superfl}.
The main feature is the onset of parametric oscillation for pump
intensities $I_p$ within a certain range of values.
Outside this range, the pump-only solution $S=I=0$ remains a
dynamically stable solution of the mean-field dynamics.

While the amplitude $P$ of the pump mode is completely determined (both
in modulus and in phase) by the incident laser amplitude, the signal
and idler phases remain free thanks to the
invariance of the equation of motion \eq{eq_mot} and of the
Hamiltonian \eq{par_ham} under a simultaneous phase rotation of the signal
and idler in opposite directions:
\begin{equation}
S\rightarrow S\, e^{i\Delta \phi} \hspace{1cm} I\rightarrow I\, e^{-i\Delta \phi}.
\label{si-sym}
\end{equation}
In the parametrically oscillating state, $S$ and $I$ have a
non-vanishing value, so that this $U(1)$ phase-rotation symmetry is
spontaneously broken.
In the absence of external perturbations, the specific value of the
phase of $S$ (and consequently of $I$) is randomly selected at each
realization of the experiment.
Consequences of the underlying symmetry are however present: no
restoring force opposes a simultaneous and opposite
rotation of the signal and idler phases, which then slowly diffuse in
time under the effect of fluctuations~\cite{phase_diffusion}. 
In the next section, we shall show how the spontaneosly broken
symmetry manifests itself as a strong response of the system when
spatial twists of the signal/idler phases are created by an extra {\em
  probe} laser at frequency
$\omega_r=\omega_s+\Delta\omega$ and wavevector $k_r=k_s+\Delta k$
close to the frequency and wavevector of the signal emission $|\Delta
k|\ll|k_s-k_p|$.

\section{Response to a weak perturbation and the Goldstone mode \label{sec:lin}}

Provided the applied perturbation is weak, the response of the polariton
field can be calculated by linearising the mean-field equations of
motion around the solution \eq{psi0}. 
Modifying this solution as  
\begin{eqnarray}
S &\rightarrow& S + u_s\, e^{i(\Delta k\, x-\Delta\omega\, t)}
+v^*_s\, e^{-i(\Delta k\, x-\Delta \omega\, t)}\\
P &\rightarrow& P + u_p\, e^{i(\Delta k\, x-\Delta\omega\, t)}
+v^*_p\, e^{-i(\Delta k\, x-\Delta \omega\, t)}\\
I &\rightarrow& I + u_i\, e^{i(\Delta k\, x-\Delta\omega \,t)}
+v^*_i\, e^{-i(\Delta k\, x-\Delta \omega\, t)},
\end{eqnarray}
the deviations from the steady state can be grouped in a 6-vector 
$\mathcal U=(u_s,u_p,u_i,v_s,v_p,v_i)^T$ that obeys the equation
\be
\Delta \omega\, {\mathcal U}-{\mathcal L}(\Delta k) \,{\mathcal U} =
{\mathcal F}_r
\label{eqlin}
\ee
with a force vector which for our specific excitation scheme reads
$\mathcal F_r=(F_r,0,\ldots,0)^T$. 
The observable quantity is the number of polaritons created in the
$k_r$ mode, which corresponds to the square modulus of the first element
$u_s$ of the system response ${\mathcal U}$. 

The matrix $\mathcal L$ has the typical Bogoliubov structure
\be
\mathcal{L}(\Delta k) =\left( 
\begin{array}{cc}
M(\Delta k) & Q(\Delta k) \\ 
-Q^{\ast }(\Delta k) & -M^*(-\Delta k) 
\end{array}%
\right),
\label{structL}
\ee
where the $3\times 3$ matrices $M(k)$ and $Q(k)$ are defined as:
\begin{eqnarray}
M_{mn}(k)&=&[ \varepsilon(k_m+k)-\omega_m-i\gamma/2]\,\delta_{m,n} \nonumber \\ 
&+&2  \sum_{rt=1}^3\, g_{k_m+k,k_n+k,k_t} \delta_{m,n+r-t} A^*_r A_t \\
Q_{mn}(k)&=&  \sum_{rt=1}^3 g_{k_m+k,k_r,k_s} \delta_{m+n,r+t}\, 
A_r A_s. 
\end{eqnarray}
Here we have identified the three modes $s,p,i$ with respectively the values
$1,2,3$ of the matrix and vector indices, so that e.g.  
$A_1=S$, $A_2=P$ and $A_3=I$.

\begin{figure}[tb]
\begin{center}
\includegraphics[width=\columnwidth,angle=0,clip]{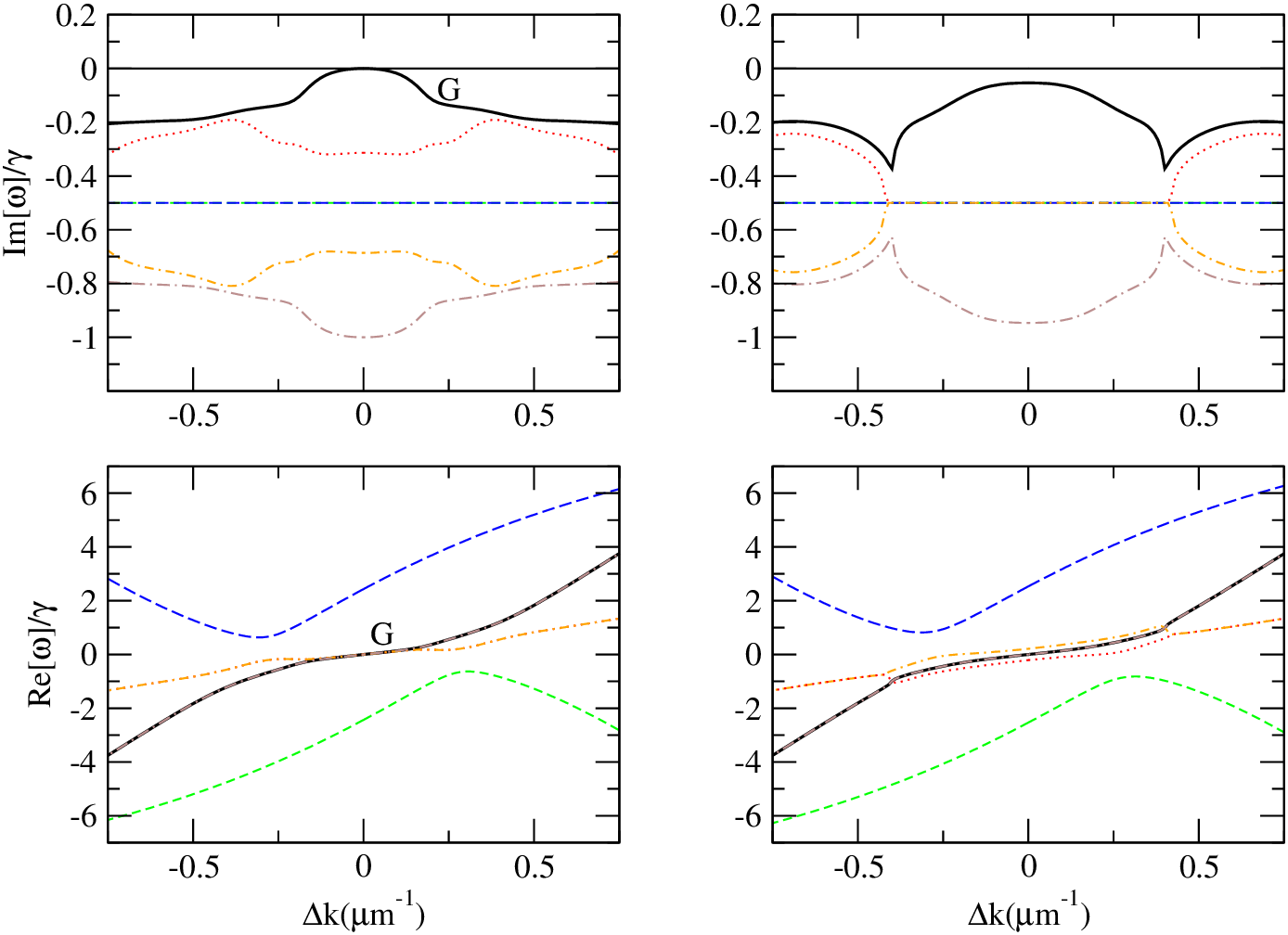}
\caption{
Imaginary (upper panels) and real (lower panels) part of the
excitation spectrum around the stationary state for the pump
intensity corresponding to the dashed line in fig.\ref{fig:sp}.
The left panels refer to the case in which the signal/idler phase
rotation symmetry is spontaneously broken and a Goldstone mode $G$
is present (heavy line). 
The right panels refer to the case where a small signal laser beam
($I_s/I_p=0.0023$) is present to explicitely break this symmetry
and destroy the Goldstone mode.
 Cavity and pump parameters as
in figure \ref{fig:sp}}
\label{fig:gs0}
\end{center}
\end{figure} 

The real and imaginary parts of the eigenvalues of the Bogoliubov
matrix $\mathcal L(\Delta k)$ give respectively the frequency and
linewidths of the elementary excitations around the parametrically
oscillating stationary state.
These fix the shape of the luminescence peaks as observed e.g. in the
experiment of~\cite{ciuti-offbranch}, as well as the poles of the
response to the probe beam.
A typical example of this Bogoliubov dispersion as a function of
$\Delta k$ is plotted in the left panels of Fig. \ref{fig:gs0}:
the most relevant feature is the presence of a branch of eigenvalues
$\omega_G(\Delta k$) tending to exactly zero for $\Delta k\rightarrow
0$.

The presence of this branch is a direct consequence of the fact that the mean
field steady state \eq{psi0} spontaneously breaks the $U(1)$ symmetry of
the signal/idler phase: as any global phase rotation of $S,I$ of the form
\eq{si-sym} maps a stationary state of \eq{eq_mot} into another
stationary state with different signal/idler phases, the generator
$G^T=(iS,0,-iI,-iS^*,0,iI^*)$ of the signal/idler phase rotations 
is an eigenvector of the matrix ${\mathcal L}(\Delta k=0)$ with zero eigenvalue
$\omega_G(\Delta k=0)=0$. 
For finite values of $\Delta k$, the soft Goldstone branch
at $\omega_G(\Delta k)$ corresponds to a spatially varying twist of
the signal/idler phases.

The existence of a soft Goldstone mode is a general result valid for
both non-equilibrium~\cite{cross}  and
equilibrium~\cite{goldstone-stat-mech} systems: in this latter case,
the Goldstone mode 
corresponds to the magnon mode of ferromagnets~\cite{magnon}, or to
the zero sound  
mode in superfluid liquid Helium and Bose-Einstein
condensates~\cite{forster,pines-nozieres}.

Fundamental differences however exist, which make the physics of
the two cases quite distinct.
In Bose systems at equilibrium, the dispersion of the Golstone 
mode goes as $\re[\omega(k)]\simeq
c_s\,|k|$ around $k=0$ and shows a singularity at $k=0$. 
This branch corresponds to weakly damped
\footnote{The damping of sound in Bose gases in the collisional
regime has a $\im[\omega(k)]\propto k^2$ dependence on wavevector,
which is however to be contrasted to the $\im[\omega(k)]\propto|k|$
dependance of a collisionless Bose-Einstein condensate in the
collisionless regime~\cite{sandro-lev}.}
sound waves which in the long-wavelength limit propagate at the speed
of sound $c_s$. 

In the present non-equilibrium case, no singularity appears in the
dispersion relation $\omega_G(\Delta k)$ of the Goldstone mode.
In particular, the real part $\re[\omega(\Delta
  k)]\propto \Delta k$ has a continuous and non-vanishing slope at
$\Delta k=0$. This fact is due to the flow of the 
pump polaritons which are injected with a finite wave vector $k_p$
  and are able to drag the elementary excitations.
On the other hand, convective stability and analyticity arguments (see Appendix
\ref{app:analytic}) show that the imaginary part goes as
$\im[\omega(\Delta k)]\approx -\alpha\,(\Delta k)^2$ for small $\Delta
k$  with a positive $\alpha>0$.

Combining these facts together, one can physically summarize that
the Goldstone mode of a planar OPO consists of a spatially slowly varying twist of
the signal and idler phases, which however does not propagate as a sound 
wave, but has a diffusive character.
Once a localized perturbation is created in the system, this will
simply relax back to the equilibrium state while it is dragged by the
pump polariton flow.
This overdamped character is a remarkable difference with respect to
the equilibrium case.

\begin{figure}[tb]
\begin{center}
\includegraphics[width=\columnwidth,angle=0,clip]{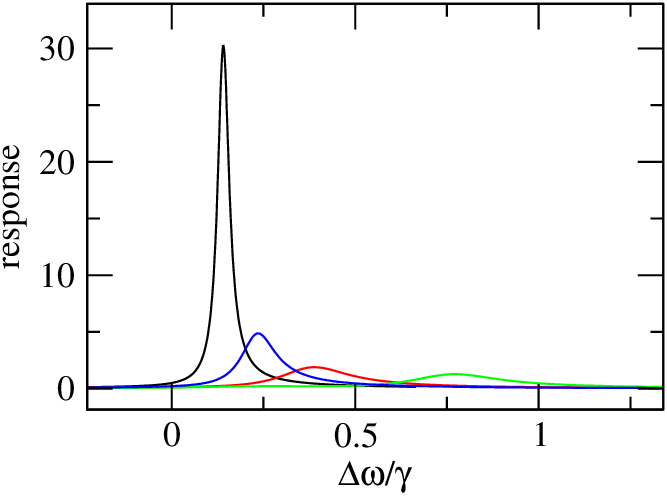}
\caption{
Response spectrum to a probe beam for several values of the wavevector
$\Delta k=k_{r}-k_s=0.1,\, 0.15, \, 0.2, \,0.3 \,\mu\textrm{m}^{-1}$
(black, blue, red, green). Cavity and pump parameters are the same as
in figure \ref{fig:sp}. 
}
\label{fig:is0}
\end{center}
\end{figure} 

The general properties of the eigenvalues of the Bogoliubov matrix
\eq{structL} can be used to physically understand the response
spectra shown in Fig.\ref{fig:is0}: the number of polaritons created
by the probe in the $k_r=k_s+\Delta k$ mode is here plotted as a
function of its frequency $\omega_r=\omega_s+\Delta \omega$.
The most apparent feature is the strong and narrow peak that appears
at low frequencies for a probe wavevector $k_r$ close to the signal
one $k_s$. In particular, its linewidth can be much smaller than the
damping rate $\gamma$ of the polariton mode: as $k_r$ approaches
$k_s$, the linewidth goes to $0$ and the peak height diverges, while
the peak dramatically broadens for increasing $\Delta k$.
Comparing these spectra with the dispersion of the elementary
excitations shown in Fig.\ref{fig:gs0}, one can immediately see that
the strong peak corresponds to the pole at $\omega_G(\Delta k)$, and
therefore to the excitation of the soft Goldstone mode.

An experimental observation of this narrow peak in the probe
response spectra with the characteristic dependence on the probe
angle would provide a unambiguous signature of the Golstone mode.
In our previous work~\cite{coherence}, the excitation of the Goldstone
mode by the quantum fluctuations showed up as a
$k$-space broadening of the signal emission in the absence of
any probe.
Although this broadening of the luminescence pattern is of conceptual
interest from the point of view of low-dimensional non-equilibrium
physics, the pump-probe experiment discussed here appears as more
favourable in view of the experimental observation of the Goldstone mode.
The response to the probe can be measured by comparing pairs of
spectra taken respectively in the presence and in the absence of the
probe, so to subtract out the background of signal light scattered by
the defects of the sample.

\section{Destroying the Goldstone Mode \label{sec:destroy}}

It is a well-known fact that the rotational symmetry can be
explicitely broken in a ferromagnet by adding an external magnetic
field that fixes a preferential orientation for the
magnetization and therefore opens a gap in the magnon
  dispersion~\cite{magnon}.
While a pinning of the Bose field phase is hardly obtained in Helium
or atomic Bose-Einstein condensates because of particle number
conservation~\cite{goldstone-stat-mech}, it can be easily done in the optical
  experiment proposed here by applying an extra laser beam (called
hereafter {\em signal laser}) at exactly the signal frequency $\omega_s$ and
wave vector $k_s$~\footnote{Note that the beam playing the role of our
  {\em signal beam} was called in~\cite{ciuti-offbranch} the {\em
    probe beam}. We have been forced to choose a different terminology
  because a {\em probe beam} has already appeared in our discussion,
  i.e. the beam probing the Bogoliubov spectrum.}.
In this way, stimulated processes push the parametric emission
to preferentially occur with a phase pinned to the incident signal
laser one.
This pinning of the signal field phase is the parametric and multi-mode
analog of what happens in a single-mode laser when a coherent signal is 
injected in the cavity~\cite{degiorgio,scully}.
In the pattern formation literature, the idea of forcing a specific
pattern shape by means of an additional beam often goes under the name
of {\em pattern synchronization}.
First studies of such issues in an optical context have recently appeared
using a completely different setup~\cite{Neubecker}.

\begin{figure}[tb]
\begin{center}
\includegraphics[width=\columnwidth,angle=0,clip]{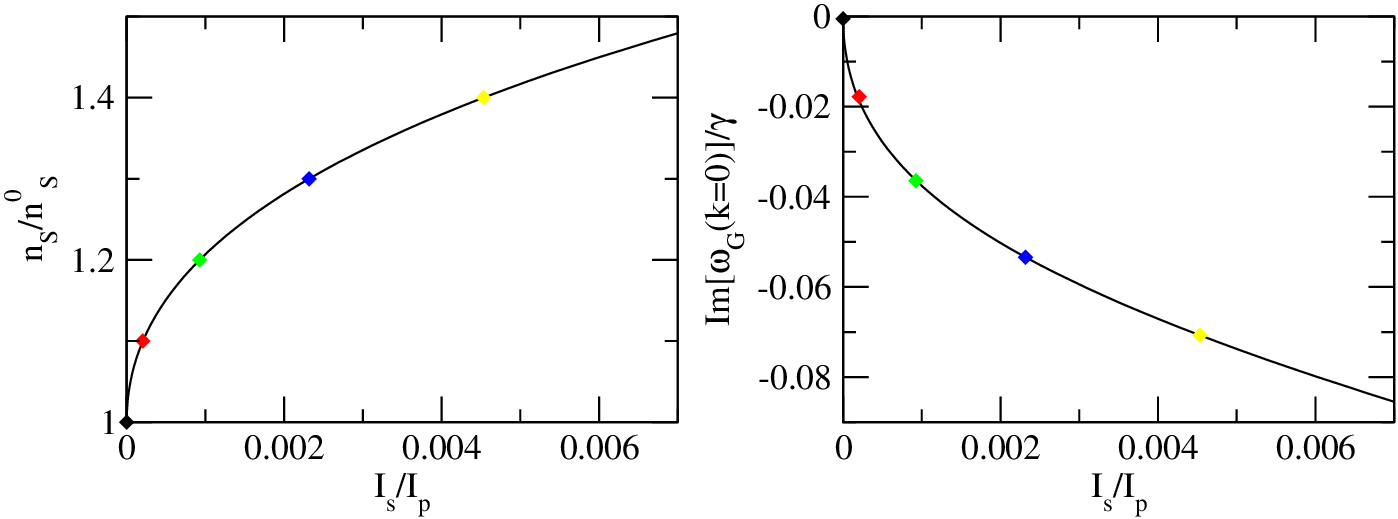}
\caption{
Signal mode occupation (left panel) and amplitude of the gap
$\im[\omega_G(\Delta k=0)]$ (right
panel)  as a function of the 
signal laser power $I_s$. The coloured points mark the $I_s$ values
of the spectra in Fig.\ref{fig:absorb1}.
}
\label{fig:Fs1}
\end{center}
\end{figure} 

As the Goldstone mode has a pole at exactly the signal laser frequency
and wavevector, the change of $S$ due to the application of the signal
beam cannot be calculated by means of the linearized theory, but has
to be calculated including he signal laser beam as an additional term
in the full equation of motion \eq{eq_mot}, as shown in Appendix
\ref{app:calculation}.

The left panel of Fig. \ref{fig:Fs1} shows the signal amplitude as a
function of the signal laser amplitude: for small values of $I_s$, 
the signal intensity $|S|^2$ shows a square root behaviour as a function of $I_s$.
As $S$ is an analytic function of $F_s$, one can expand $S \simeq
S_0+c F_s$, the square modulus of which immediately gives the
square-root behaviour $|S|^2-|S_0|^2 \simeq c \,|S_0|\,|F_s|\propto
\sqrt{I_s}$.

The effect of the signal laser on the Bogoliubov dispersion is shown
in the right panels of of Fig. \ref{fig:gs0}:
the imaginary part  $\im[\omega_G(\Delta k)]$ shows now a finite gap
at $\Delta k=0$, so that no Goldstone mode exist any longer.
The damping rate of Bogoliubov excitations is finite for any value of
$\Delta k$.
On the other hand the real part is much less affected by the signal
laser beam and, in particular, remains $0$ at  $\Delta k=0$.

\begin{figure}[tb]
\begin{center}
\includegraphics[width=\columnwidth,angle=0,clip]{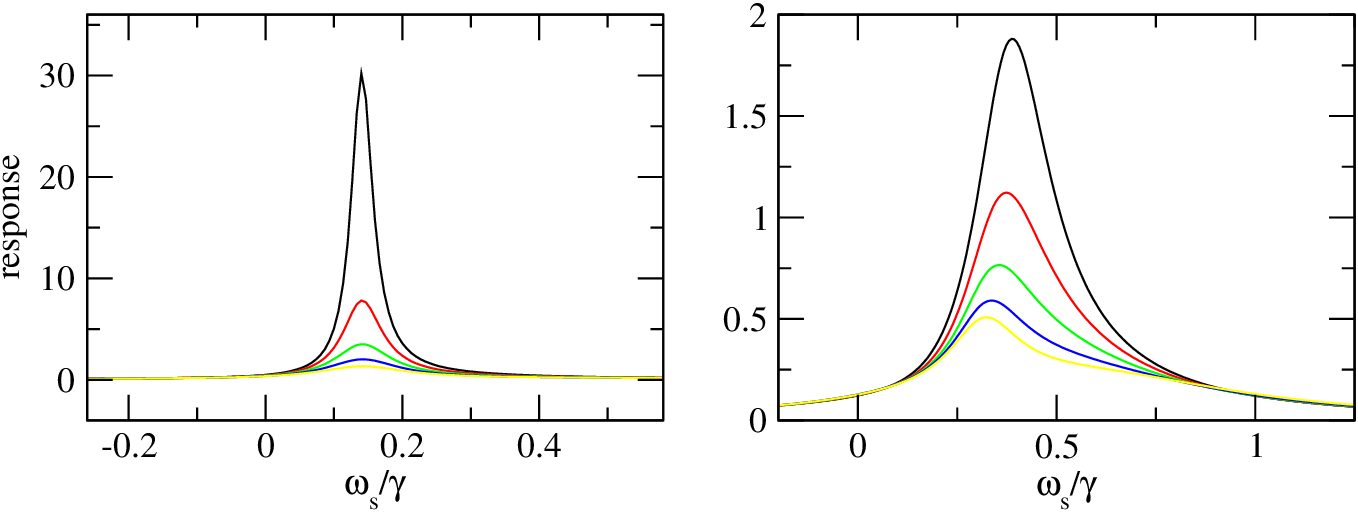}
\caption{
Response spectrum of the polariton gas to an extra laser beam at $\Delta k=0.1\,\mume$
(left panel) and $\Delta k=0.2\,\mume$ (right panel) for several values of
the signal laser intensity $I_s$ marked with the symbols in Fig.\ref{fig:Fs1}. 
}
\label{fig:absorb1}
\end{center}
\end{figure} 

This is another important difference as compared to the
thermodynamical equilibrium case, where the presence of the
external field explicitely breaking the symmetry opens a gap in the
real part of the dispersion law~\cite{magnon}. 
In the present non-equilibrium case the gap appears instead in the
imaginary part. 
The dependence of the gap amplitude $\im[\omega_G(\Delta k=0)]$ on the
signal laser intensity $I_s=|F_s|^2$ is plotted in the right panel of
Fig. \ref{fig:Fs1}.  
Again, the behaviour for low
values of $I_s$  is a square root one. This can be easily explained by
the fact that $\omega_G(\Delta k=0)$ is an analytic function of $S,P$
and $I$ and these have the square root dependence on $I_s$ shown in
the left panel of Fig.\ref{fig:Fs1}.

The presence of the gap in the Bogoliubov spectrum is expected to have
dramatic 
consequences on the response of the system to the probe laser of
wave vector $k_r$ and frequency $\omega_r$.
This can be quantitatively studied by means of the same linear
response equation \eq{eqlin} once the correction to the stationary-state
amplitudes $S$, $P$ and $I$ due to the presence of the signal laser
have been duly taken into account.
Note how this merely quantitative change in the entries of the
${\mathcal L}$ matrix is enough to destroy the weakly-damped Goldstone
branch.

Examples of probe response spectra as a function of the probe frequency
$\omega_r$ are plotted in Fig. \ref{fig:absorb1} for several values of
the signal laser power $I_s$ and two different values of the probe
wave vector $k_r$.
It is easy to see that the main consequence of the presence of the
signal laser field is to broaden the peak and suppress the strongly
peaked response to small $\Delta k$ perturbations.
This is in exact correspondance with the Bogoliubov spectra plotted in
the right panels of Fig.\ref{fig:gs0}.

This characteristic dependence of the probe response on the
 signal beam intensity constitutes a direct and experimentally
 accessible signature of
 the existence of a Goldstone mode which gets destroyed when the 
 signal/idler phases are pinned by the signal beam.

\section{Signal frequency mismatch \label{sec:mismatch}}

The discussion of the previous section has assumed the signal laser
field to be exactly on resonance with the natural parametric
oscillation frequency $\omega_s$ of the cavity when this is
illuminated by the pump laser alone. 
Here we shall study the case when the parametric emission is
forced by the signal beam to take place at a slightly different
frequency $\omega_s+\delta \omega_s$.

\begin{figure}[tb]
\begin{center}
\includegraphics[width=\columnwidth,angle=0,clip]{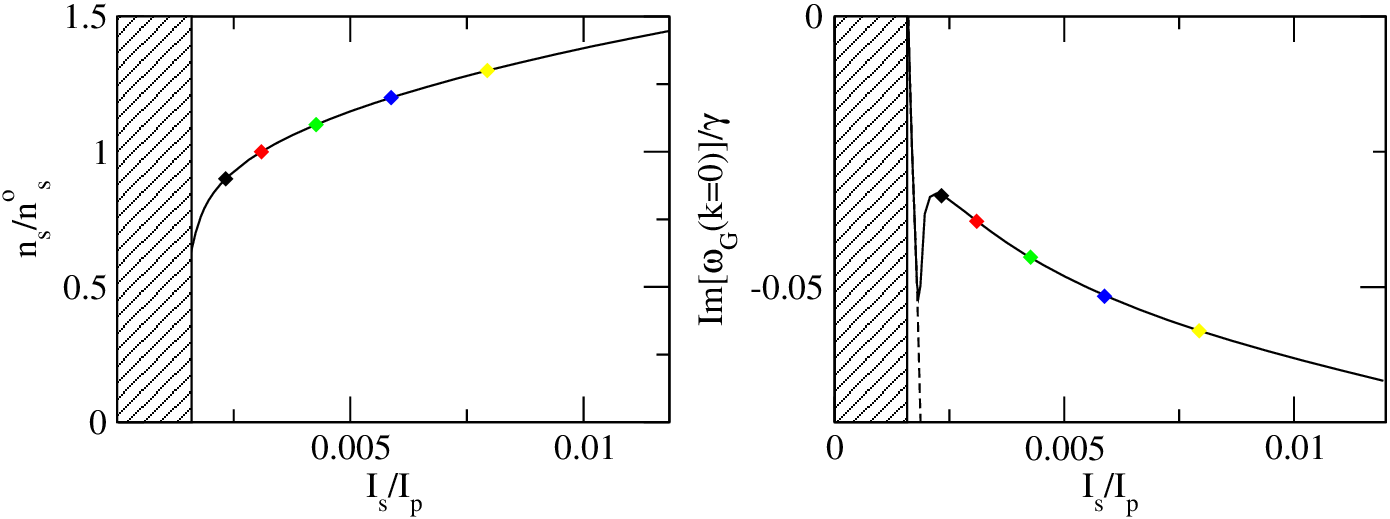}
\caption{
The same as figure \ref{fig:Fs1}, but for a signal laser detuning of
$\delta\omega_s=0.01\,\textrm{meV}$.
In the hatched region, the three-mode solution is dynamically unstable
(see text). 
}
\label{fig:Fs2}
\end{center}
\end{figure} 

Two intensity regimes have now to be distinguished for the signal laser. 
For a small intensity $I_s$, the signal beam can be considered
as a perturbation of the parametric oscillation state at the natural
frequency $\omega_s$ and can be introduced within the linear response
theory of section \ref{sec:lin} as a probe beam.  
Thanks to the detuning $\delta\omega_s$, the signal beam is in
fact not on resonance with the Goldstone mode, so that its response
does no longer diverge as it instead happened in sec.\ref{sec:destroy}.

On the other hand, the linearized theory breaks down for larger signal
laser intensity, when the parametric oscillation no longer takes
place at its natural  frequency $\omega_s$ but rather
appears at the forcing signal beam frequency $\omega_s+\delta\omega_s$.
A stationary state for the mean-field equation \eq{eq_mot} of the form:
\begin{multline}
\psi_{LP}(x,t) =P\, e^{i(k_p x-\omega_{p}t)}+
S\, e^{i(k_s x-(\omega_{s}+\delta\omega_s)t)}+\\
+I\, e^{i(k_i x- (\omega_i-\delta\omega_s) t)},
\label{psi0-2}
\end{multline}%
has thus to be considered.
The transition from one case to the other can be described by means of
a Hopf bifurcation scenario in a theory explicitely including the
possibility of a time-dependence for $S$, $P$ and $I$.
For the sake of simplicity, we will restrict ourselves in the following to the
parameter range well above the bifurcation point, where the ansatz
\eq{psi0-2} is accurate. 
The Hopf bifurcation point is signalled by the solution \eq{psi0-2}
becoming unstable, i.e. the imaginary part of an eigenvalue of the 
linearized theory becoming positive $\im[\omega(\Delta k=0)]>0$.

\begin{figure}[tb]
\begin{center}
\includegraphics[width=\columnwidth,angle=0,clip]{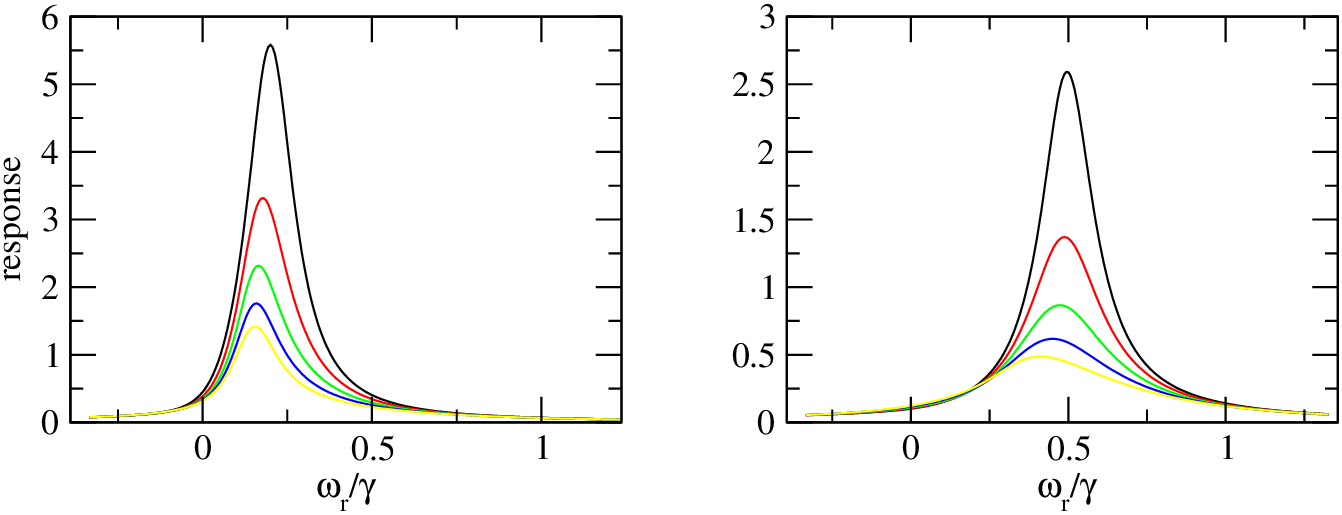}
\caption{
The same as Fig. \ref{fig:absorb1}, but for a signal laser detuning of
$\delta \omega_s=0.01{\textrm meV}$ and
for the values of $I_s$ indicated with the symbols in Fig.\ref{fig:Fs2}.
}
\label{fig:absorb3}
\end{center}
\end{figure} 

The signal mode occupation $n_s$ and the imaginary part
$\im[\omega_G(\Delta k=0)]$ of
the Goldstone mode at $\Delta k=0$ are plotted in respectively
the left and right panels of Fig. \ref{fig:Fs2} as a function of the
signal beam intensity $I_s$.
The hatched region is the low $I_s$ region where the solution
\eq{psi0-2} is dynamically unstable.
Well above the instability region, the behaviour of
$\im[\omega_G(\Delta k=0)]$ as a function of the signal laser
intensity recovers the behaviour of the zero-detuning
$\delta\omega_s=0$ case.

Examples of response spectra to the probe beam are shown in
Fig.\ref{fig:absorb3}.
For low values of the signal laser intensity $I_s$ (but still above
the instability threshold) the response is strongly peaked at the
weakly damped Goldstone mode, while it broadens for larger values of
$I_s$: this phenomenology is in close qualitative agreement with
the one shown in Fig. \ref{fig:Fs1} for the case of a perfectly tuned
signal laser.
Provided we are sufficiently far from the Hopf bifurcation, a mismatch
in the signal frequency therefore does not substantially affect the
signal/idler phase pinning effect which is responsible for the
disappearance of the strong response associated to the Goldstone mode.

\section{Finite spot \label{sec:finite}}

In all the previous discussion, a spatially homogeneous system was
considered, with a spatially homogeneous pump beam; in that case, the
wave vector was a good quantum number.
Real experiments are however performed using finite-size pump laser
spots, generally of Gaussian shape
$F_p(k) \sim \exp[-\sigma_p^2\,(k-k_p)^2/2]$.
This implies that a range of wave vectors are excited, yet the
 pump, signal and idler beams remain well distinct in $k$ space provided the
(real-space) spot size $\sigma_p$ is large enough $| k_p-k_s|\,\sigma_p\gg 1$.
Under such an hypothesis, the signal and idler phase rotation
\eq{si-sym} is still a symmetry element of the problem
\footnote{Rigorously speaking, simultaneous and opposite rotations
of the signal/idler phases without affecting the pump one requires
a boundary to be defined separating them.
This can be done without breaking the wave function's continuity in
$k$ space only if the polariton population is negligible in between. 
This is a reasonable assumption if $\sigma_p\gg |k_s-k_p|^{-1}$,
that is if the characteristic length scale of the spot profile is much
wider than the wavelength of the fringe pattern created by the
interference between signal, pump and idler.
The physics is very different in the case of B\'enard cells in a box
geometry with sharp boundaries, as pointed out in~\cite{sneddon}.
In this case, the displacement of the roll pattern has to be
accompanied by a deformation of 
the cells close to the boundaries, so that a sort of restoring force
appears which opposes to the displacement.
Within our formalism, the presence of this symmetry-breaking restoring
force appears as all modes of the Bogoliubov spectrum being damped.
In our case, the boundaries are smooth and the pattern wavelength
is much shorter than the characteristic thickness of the boundary
layer, so that the pattern can shift almost freely through the spot.
This means that the symmetry-breaking restoring force is small and can
generally be neglected.
}.

\begin{figure}[tb]
\begin{center}
\includegraphics[width=\columnwidth,angle=0,clip]{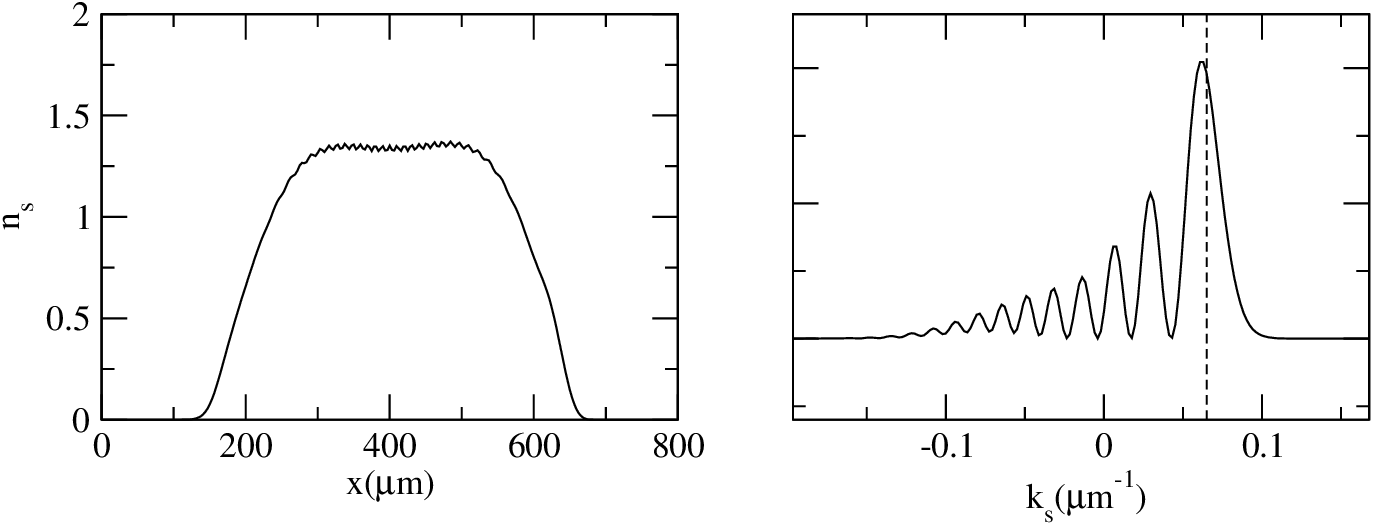}
\caption{
Numerical results for the spatial profile of the signal density (left
panel) and the signal wave vector spectrum (right panel) for a pump
waist $\sigma_p=300\,\mume$, a carrier pump wave vector
$k_p=1.57\,\mume$. The vertical dashed line in the right panel 
shows the wavevector of the signal emission at the center of the 
spot. Other parameters as in Fig.\ref{fig:sp}.
}
\label{fig:select}
\end{center}
\end{figure} 

The spatial and wave vector distributions of the parametric signal
emission are shown in Fig. \ref{fig:select} for the case of a wide pump
spot in the absence of any incident signal laser.
The plotted curves are the result of a numerical integration of the
full mean-field evolution equation \eq{eq_mot} until the steady-state is
reached:
the main feature to note is the inhomogeneous wave
vector broadening that can be observed in the right panel, a
broadening which always remains
much smaller than the wavevector space distance $|k_p-k_s|\simeq
  1.5\,\mu\textrm{m}^{-1}$. A detailed explanation of the physical
  origin of this broadening is posponed to the forthcoming
  publication~\cite{pattern}.

\begin{figure}[tb]
\begin{center}
\includegraphics[width=\columnwidth,angle=0,clip]{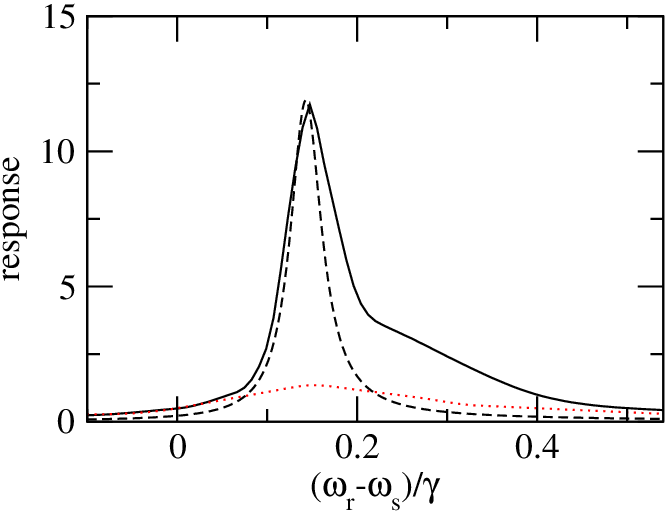}
\caption{
Solid line: response of the polariton pattern of Fig. \ref{fig:select} as a function of frequency, 
when it is probed by a laser beam with wave vector $k_{r}=0.175\mu \rm m^{-1}$
(full black line).
Dashed line: prediction for a homogeneous system with the values
at the center of the spot. 
Red dotted line shows the response when a signal laser with intensity
$I_s=10^{-3}\, I_p$ and waist $\sigma_s\gg \sigma_p$ is applied to the
inhomogeneous system.  
}
\label{fig:goldstone-fin}
\end{center}
\end{figure}

The response of this steady-state to an additional probe laser beam
with wave vector $k_{r}=0.175$ and a waist $\sigma_r=\sigma_p$ is
shown in Fig. \ref{fig:goldstone-fin}.
As in the case of an homogeneous system, the very peaked response as a
  function of frequency is a consequence of the presence of the Goldstone 
mode.
The enhanced broadening as compared to the homogeneous case is
  mostly due the inhomogeneous broadening of the 
signal emission in wavevector space shown in fig.\ref{fig:select}.
The top of the peak is in good agreement with the prediction for a
homogeneous system with the same signal wave vector as the one
actually selected at the center of the finite spot (dashed line).
Agreement is observed also in the low-frequency tail of the peak,
but differences are apparent in the high-frequency tail: the tail of
the finite spot curve extends in fact for much longer than in the
homogeneous case.
This is a direct consequence of the asymmetric inhomogeneous
broadening of the signal wave vectors shown in the right panel of
Fig. \ref{fig:select}, which indeed mostly extends towards lower $k_s$ values. 
Relative to these wave vectors, the perturbation has a larger $\Delta
k=k_r-k_s$, which means that the response peak lies at higher
frequencies and has a larger width (as shown in the lower panels of 
Fig.\ref{fig:gs0}).

The red dotted line in Fig. \ref{fig:goldstone-fin} shows the response
when a signal laser beam is applied, which explicitely breaks the
  $U(1)$ symmetry.
Analogously to the case of a homogeneous system, the fact that the
$U(1)$ symmetry is now explicitely broken causes a strong broadening of
the peak and a corresponding suppression of the peak response.
This completes the proof that the concepts and the analytical results
developed for the homogeneous case can still be safely applied  when
the pump spot has a finite size.

\section{Conclusions}
\label{sec:Conclu}
In the present paper, we have studied the properties of the Goldstone
mode that appears in the elementary excitation spectrum of 
planar parametric oscillators above threshold as a consequence of the
spontaneous breaking of the $U(1)$ signal/idler phase symmetry.
Although quantitative predictions have been shown only for the case of
planar semiconductor microcavities with a quantum well excitonic
transition strongly coupled to the cavity mode, the physics is the
same in any kind of parametric oscillator with planar geometry.

Analogies and differences with thermodymical equilibrium systems
showing the same spontaneous symmetry breaking phenomenon
(e.g. superfluid Helium 4 or Bose-Einstein condensates) have been
drawn. 
In particular, several features have been pointed out, which are 
typical of the stationary state of a non-equilibrium system where a
dynamical equilibrium occurs between driving and losses.
Although its dispersion tends as usual to $0$ in the long wavelength
limit, the Goldstone mode of our non-equilibrium system is not a
propagating mode like zero-sound in Helium or Bose-Einstein
condensates, but rather an overdamped mode. 
A simple way of probing its dispersion at small wavevectors is to
measure the response of the system to an additional probe laser shone onto the
cavity at an angle close to the signal emission.
As the probe beam approaches the signal, the damping rate of the
Goldstone mode tends to vanish, which corresponds to a probe response
spectrum extremely peaked at the Goldstone mode frequency. 

A signal laser beam injected into the cavity and stimulating the
parametric emission is shown to provide an external field which
explicitely breaks the $U(1)$ symmetry. 
This in analogy with what happens in a ferromagnet, where an external
magnetic field provides a preferential orientation to the
magnetization.   
As usual, the explicit breaking of the symmetry is responsible for the
opening of a gap in the Goldstone mode dispersion. 
Differently from the equilibrium case, the gap is here in the
imaginary part of the dispersion.
As a consequence, the peak in the response spectrum gets broader and
eventually almost disappears when the signal laser intensity grows
higher.

In the last sections of the paper, a few issues of experimental
  relevance in view of the actual observation of the Goldstone
  mode have been addressed, such as the effect of a slight detuning of
  the signal laser from the natural parametric oscillation frequency,
  and the finite size of the excitation spot. All of them are
  shown not to give any dramatic effect on 
the observation of the Goldstone mode.

An experimental study along these lines appears therefore feasible
with present technology samples and will constitute an
important step in the understanding of the properties of parametric
oscillation in spatially distributed systems, in particular of its
analogies and differences with the Bose-Einstein condensation phase
transitions in systems at thermodynamical equilibrium.  

\acknowledgments
Continuous stimulating discussions with Cristiano Ciuti, Jer\^ome
Tignon, Carole Diederichs, Herv\'e Henry, Arnaud Couairon,
Jean-Marc Chomaz, Jozef Devreese and Jacques Tempere are warmly acknowledged.
This research has been supported financially by the FWO-V projects Nos.  
G.0435.03, G.0115.06 and the Special Research Fund of the University of
Antwerp, BOF NOI UA 2004.
M.W. acknowledges financial support from the FWO-Vlaanderen in the form 
of a ``mandaat  Postdoctoraal Onderzoeker''.

\begin{appendix}

\section{Equations defining the stationary-state}
\label{app:calculation}

Substituting the ansatz \eq{psi0} in the equations of motion
\eq{eq_mot} and imposing the stationary state leads to the following
set of complex equations:
\begin{widetext}
\begin{eqnarray}
0&=&\frac{1}{X_{p}^{2}}%
\left[ \epsilon_p-i\frac{\gamma}{2}-\omega _{p}\right] \tilde{P}+\left( \vert
\tilde{P}\vert ^{2}+2\vert \tilde{S}\vert ^{2}+2\vert \tilde{I}\vert
^{2}\right) \tilde{P}+2\tilde{P}^{\ast }\tilde{S}\tilde{I}+\tilde{F}_{p}  \label{mf1} \\
0 &=&\frac{1}{X_{s}^{2}}%
\left[ \epsilon_s-i\frac{\gamma}{2}-\omega _{s}\right] \tilde{S}+\left( 2\vert
\tilde{P}\vert ^{2}+\vert \tilde{S}\vert ^{2}+2\vert \tilde{I}\vert
^{2}\right) \tilde{S}+\tilde{P}^{2}\tilde{I}^{\ast }  \label{mf2} \\
0 &=&\frac{1}{X_{i}^{2}}%
\left[ \epsilon_{i}-i\frac{\gamma}{2}-2\omega _{p}+\omega _{s}\right] \tilde{I}+\left(
2\vert \tilde{P}\vert ^{2}+2\vert \tilde{S}\vert ^{2}+\vert
\tilde{I}\vert ^{2}\right) \tilde{I}+\tilde{P}^{2}\tilde{S}^{\ast }.  \label{mf3}
\end{eqnarray}%
\end{widetext}
The following shorthand notations have been introduced
$\epsilon_{p,s,i}=\varepsilon(k_{p,s,i})$ and $X_{p,s,i}=U_X(k_{p,s,i})$.
Scaled quantities $\tilde{S}= \sqrt{g}\,X_{s}\, S$,  $\tilde{P}= \sqrt{g}\, 
X_{p}\, P$,  $\tilde{I}= \sqrt{g}\, X_{i}\, I$ and
$\tilde{F}_{p}=\sqrt{g}\,U_C(k_p)/U_X(k_{p})\,F_{p}$ have been
defined. 
The complex variables ${\tilde S}$, ${\tilde P}$, ${\tilde I}$,
$\omega_s$ can be obtained from the set of
three complex equations (\ref{mf1}-\ref{mf3}) together with the
condition that the frequency $\omega_s$ has to be real: only 7 real
equations are then available to determine 8 real variables, which
means that the solutions are grouped in 1D manifolds. 
As the set of equations (\eq{mf1}-\eq{mf3}) is symmetric under the
phase rotation \eq{si-sym}, the solution manifold is generated by the
action of \eq{si-sym} on a given solution.

The presence of the signal beam at frequency $\omega_s+\delta\omega_s$
is taken into account in the set of equations (\eq{mf1}-\eq{mf3}) by
simply adding the term
$\tilde{F}_{s}=\sqrt{g}\,U_C(k_s)/U_X(k_{s})\,F_{s}$ to the right-hand 
side of \eq{mf2} and considering $\omega_{s}\rightarrow
\omega_{s}\pm\delta \omega_s$ as a fixed quantity.
In this case, the number of unknowns is reduced to three complex
variables, which are then completely determined by the three complex
equations.

\section{Analytical remarks on the Goldstone mode}
\label{app:analytic}

In this appendix we show that (convective) stability~\cite{cross}
implies that only even powers 
are possible in the expansion of  
$\im[\omega(k)]$ around $k=0$.
The eigenvalues of a linear operator $\mathcal L(k)$ have a Taylor
expansion around every 
value of $k$, except for some special points where fractional powers
$1/p$ can occur  
(see Ref.\cite{kato}, p. 65).
If the Taylor expansion exists, 
stability requires $\im[\omega(k)]<0$ for all $k$, which implies that
only even powers of $k$ are allowed in the Taylor expansion 
of $\im[\omega(k)]$ around $k=0$.

If accidentally $k=0$ is a special point of order $p$, there are $p$ eigenvalues
which have a Puiseux series instead of a Taylor series 
\be
\lambda_h(k) = \lambda + \alpha_1 e^{2\pi ih/p} k^{1/p} + \alpha_2 e^{4\pi ih/p} k^{2/p}+
\cdots,
\ee
for $h=0,1,\ldots,p-1$. 
If $p=2$, stability requires that $\text{Im}(\alpha_m)=0$  
for any odd $m$, so that no half-integer powers are allowed in the expansion of
$\im[\lambda(\omega)]$. 
The case with $p>2$ is ruled out just because it would forcedly give an
eigenvalue with positive imaginary part. 

In summary, only even powers are allowed in the expansion of
$\im[\omega(k)]$, so that the expansion of $\im[\omega(k)]$ is to
leading order quadratic (unless this term accidentally vanishes).

\section{Comparisons with $\chi^{(2)}$ OPO's}
\label{app:chi2}
In this Appendix we make the link of our theory to the different case
of an optical parametric oscillator based on a nonlinear medium
showing a second-order optical nonlinearity instead of a third order
one~\cite{planar_OPO_2}.

In the non-degenerate case where pump, signal and idler belong to
different photonic branches of the planar cavity, the terms of the
Hamiltonian responsible for the parametric downconversion process have
the form: 
\begin{equation}
H_{\rm int}=\hbar g\,\left[\Ehd_p(k_p)\,
\Eh_i(k_i)\,\Eh_s(k_s)+{\textrm h.c.}\right],
\label{par_ham_2}
\end{equation}
where a single pump photon is converted into a pair of signal/idler photons.
An ansatz analogous to \eq{psi0} can then be used for the stationary state
of the three pump, signal and idler fields at
mean-field regime:
\begin{eqnarray}
E_p(x) &\rightarrow& P\, e^{i(k_p x - \omega_p t)} \label{ansatz2a}\\
E_s(x) &\rightarrow& S\,e^{i(k_s x -  \omega_s t)}\\
E_i(x) &\rightarrow& I\,e^{i(k_i x -  \omega_i t)},\label{ansatz2c}
\end{eqnarray}
and the amplitudes $P$, $S$, and $I$, as well as the frequency
$\omega_s$ are obtained by
inserting this ansatz into the field equations of motion for the
Hamiltonian \eq{par_ham_2}.

It is appearent that the transformation \eq{si-sym} is still a
symmetry of the Hamiltonian \eq{par_ham_2}, which is spontaneously
broken above threshold by the ansatz (\ref{ansatz2a}-\ref{ansatz2c}).
A Goldstone mode is therefore present, with the same properties as
discussed in the body of the present paper.

The only different lies in the dependence of the stationary state on
the pump intensity~\cite{lugiato_degen_OPO}: as the second-order nonlinearity does not produce
any direct shift of the mode frequency due to the mode population, the
curves of Fig.\ref{fig:sp} are dramatically modified.
In particular, there is no upper threshold for the parametric process,
which at high intensities is replaced by chaotic
behaviour~\cite{lugiato_degen_OPO}.

In the degenerate case where signal and idler belong to the same
photonic $L$ branch, the parametric terms of the parametric
Hamiltonian read instead: 
\begin{equation}
H_{\rm int}=\hbar g\,\left[\Ehd_p(k_p)\, 
\Eh_L(k_i)\,\Eh_L(k_s)+{\textrm h.c.}\right]
\label{par_ham_2_d}
\end{equation}
so that the ansatz 
\begin{eqnarray}
\psi_p(x) &\rightarrow& P\, e^{i (k_p x - \omega_p t)}\\
\psi_L(x) &\rightarrow& S\, e^{i (k_s x - \omega_s t)} + I\, e^{i (k_i x
  - \omega_i t)}  \label{psiL} 
\end{eqnarray}
has to be used to describe the parametric oscillation state.
Also in this case, the equations of motion are invariant with respect
to the transformation \eq{si-sym} and all the physics of the Goldstone
mode remains unchanged. Note how the transformation
\eq{si-sym} corresponds in this case to a spatial translation of the
field \eq{psiL} in the $L$ mode.

The only exception occurs when the system oscillates in a
completely degenerate regime $k_s=k_i=k_p/2$ where the signal and the
idler coincide: in this case, the $U(1)$ symmetry is reduced to a
discrete $\pm$ symmetry and no Goldstone mode is any longer present.
A linear analysis for this completely degenerate case has been
worked out in \cite{lugiato_degen_OPO}, where it was shown explicitely
that no zero eigenvalue is present for $\Delta k=0$. 

Note that the condition characterizing a complete degeneracy regime is
weakened for a finite 
spot of size $\sigma_p$, where the pump, signal and idler spots have a
$k$-space broadening of the order of $1/\sigma_p$.
In this case, in fact, the Goldstone mode can be shown to disappear as
soon as $\sigma_p|k_s-k_i| \simeq 1$.

\end{appendix}

\end{document}